\documentclass[aps,prd,showpacs,floatfix,nofootinbib,twocolumn]{revtex4-1}
\usepackage[dvipdfm]{graphicx}
\usepackage{amsmath,amssymb}

\begin{document}
\title{Anisotropic mass ejection from black hole-neutron star binaries:
Diversity of electromagnetic counterparts}
\author{Koutarou Kyutoku$^1$, Kunihito Ioka$^2$, Masaru Shibata$^3$}
\affiliation{$^1$Department of Physics, University of
Wisconsin-Milwaukee, P.O. Box 413, Milwaukee, Wisconsin 53201, USA\\
$^2$Theory Center, Institute of Particles and Nuclear Studies, KEK,
Tsukuba 305-0801, Japan\\
$^3$Yukawa Institute for Theoretical Physics, Kyoto University, Kyoto
606-8502, Japan\\
}
\date{\today}

\begin{abstract}
 The merger of black hole-neutron star binaries can eject substantial
 material with the mass $\sim 0.01$--$0.1 M_\odot$ when the neutron star
 is disrupted prior to the merger. The ejecta shows significant
 anisotropy, and travels in a particular direction with the bulk
 velocity $\sim 0.2 c$. This is drastically different from the binary
 neutron star merger, for which ejecta is nearly isotropic. Anisotropic
 ejecta brings electromagnetic-counterpart diversity which is unique to
 black hole-neutron star binaries, such as viewing-angle dependence,
 polarization, and proper motion. The kick velocity of the black hole,
 gravitational-wave memory emission, and cosmic-ray acceleration are
 also discussed.
\end{abstract}

\maketitle

\section{Introduction}

Black hole (BH)-neutron star (NS) binary coalescences are among the
prime sources of gravitational waves (GWs) for ground-based detectors,
such as Advanced LIGO, Advanced Virgo, and KAGRA
\cite{ligo2010,virgo2011,lcgt2010}. Gravitaional waves from BH-NS
binaries will enable us to probe the supranuclear-density matter
\cite{lackey_ksbf2012,lackey_ksbf2013} and cosmological expansion
\cite{messenger_read2012} via the NS tidal effect even without
electromagnetic (EM) observation. The imprint of the NS tidal effect is
the most prominent when the NS is disrupted outside the BH innermost
stable circular orbit \cite{kyutoku_st2010,kyutoku_ost2011}. If tidal
disruption occurs outside it, a hot and massive remnant disk may be
formed around a remnant BH. Such a system could drive an
ultra-relativistic jet; therefore, BH-NS binaries are also important as
possible progenitors of short-hard gamma-ray bursts (GRBs)
\cite{narayan_pp1992}.

Mass ejection from the BH-NS merger also has a long history of
theoretical investigation \cite{lattimer_schramm1974}, and has recently
been getting a lot more attention. One of the most important reasons is
the pressing necessity to understand EM radiation from, or EM
counterparts to, the BH-NS merger. A variety of EM counterparts will be
accompanied by the mass ejection from the merger of binary compact
objects including NSs
\cite{metzger_berger2012,piran_nr2013}. Simultaneous detection of EM
radiation with GWs are indispensable for confident GW detection and
accurate localization of the GW sources \cite{nissanke_ma2013}.

In this paper, we explore possible signatures of mass ejection from the
BH-NS merger based on our recent numerical-relativity
simulations. Numerical relativity is a unique tool to understand the
merger of binary compact objects. The numerical relativity community has
been focusing primarily on GW emission and disk formation for the BH-NS
merger (see \cite{shibata_taniguchi2011} for reviews) and is just
beginning to explore mass ejection from the BH-NS merger
\cite{kyutoku_ost2011,foucart_ddkmopsst2013,lovelace_dfkpss2013,deaton_2013}. While
characteristic properties of ejecta such as the mass and energy are
reported in these works, observational implication of the ejecta is not
fully understood yet. In particular, ejecta from the BH-NS merger shows
significant anisotropy, and thus EM counterparts could show significant
differences from ones by nearly isotropic ejecta from the NS-NS merger
\cite{hotokezaka_kkosst2013,bauswein_gj2013}.

\section{Simulation}

Our BH-NS models are chosen so that the mass ratio, BH spin, and NS
equation of state (EOS) are systematically varied. We fix the NS mass,
$M_\mathrm{NS}$, to be a representative value $1.35 M_\odot$ of observed
NS-NS binaries \cite{ozel_pnv2012}. The mass ratio of the BH to NS, $Q$,
is varied from 3 to 7, where $Q \gtrsim 5$ may correspond to a typical
BH mass in low-mass x-ray binaries \cite{ozel_pnm2010}. The
nondimensional spin parameter of the BH (the spin divided by the mass
squared), $\chi$, is chosen between 0 and 0.75. In this study, we focus
on cases in which the BH spin is aligned with the orbital angular
momentum. The NS EOS are modeled by the same piecewise polytropes as
those adopted in \cite{hotokezaka_kkosst2013}, i.e, APR4, ALF2, H4, and
MS1 (see also \cite{read_lof2009}). For these EOSs, radii of a $1.35
M_\odot$ NS span a wide range. Specifically, APR4, ALF2, H4, and MS1
give 11.1km, 12.4km, 13.6km, and 14.4km, respectively. Methods for
computing initial conditions are described in
\cite{kyutoku_st2009,kyutoku_ost2011}.

\begin{figure}[tbp]
 \begin{tabular}{c}
  \includegraphics[width=88mm,clip]{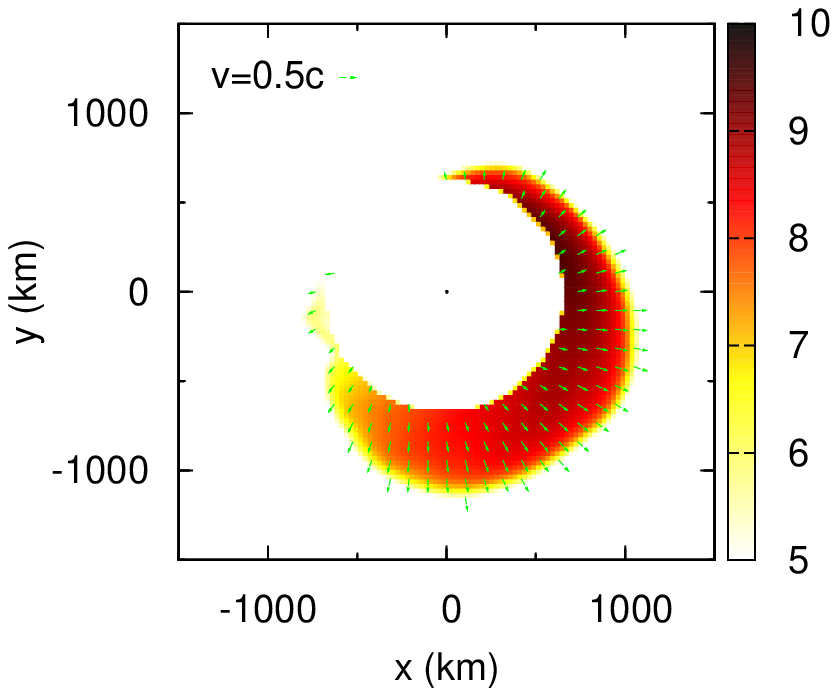} \\
  \includegraphics[width=88mm,clip]{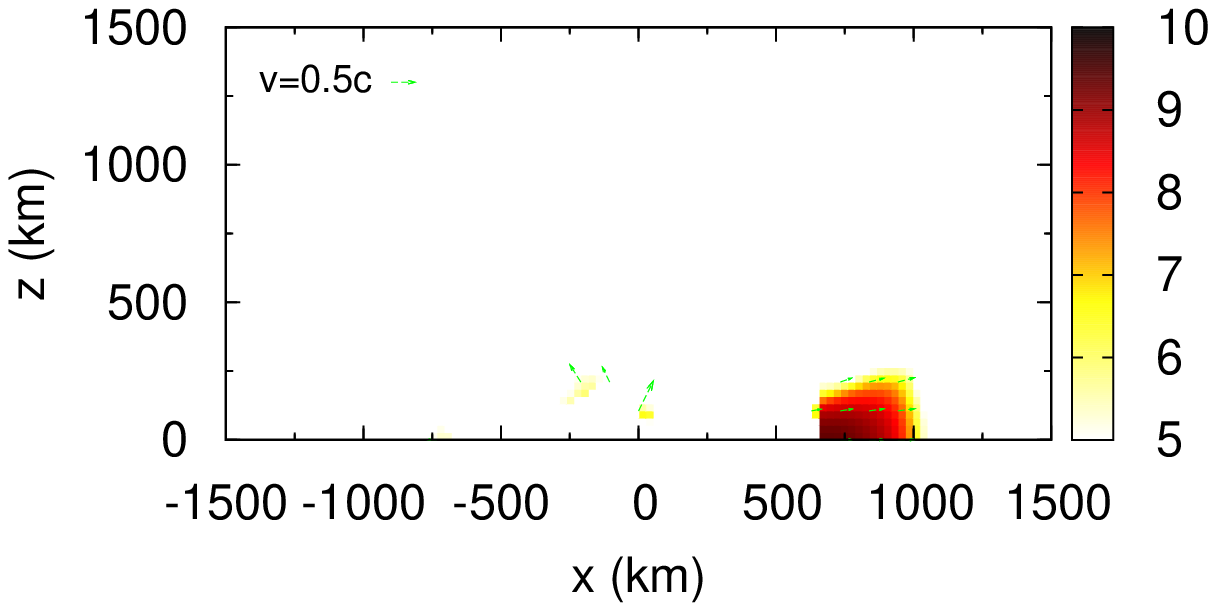}
 \end{tabular} 
 \caption{The rest-mass density ($\rho$) profiles of ejecta overplotted
 with the velocity at $\approx 10$ ms after the merger for $Q=5$,
 $\chi=0.75$, and H4 EOS model. We only show unbound material to
 elucidate geometry of the ejecta, and the blank region between the
 ejecta and BH is filled with unshown bound material. The top and bottom
 panels are for the {\it xy}- and {\it xz}-planes, respectively. The
 color panel on the right of each plot is $\log_{10} \rho$ in $\mathrm{g
 \; cm^{-3}}$. The region above $\sim 500$ km in the bottom panel is
 much clearer than that in a typical NS-NS merger (see the corresponding
 panels of Figs.~3-5 in \cite{hotokezaka_kkosst2013}).}
 \label{fig:ejmer}
\end{figure}

Our simulations are performed in full general relativity with an
adaptive-mesh-refinement code, {\tt SACRA}
\cite{yamamoto_st2008}. Improvements to previous simulations are
summarized as follows (see also \cite{hotokezaka_kkosst2013}). First, we
extend computational domains so that we can track long-term evolution of
ejecta. Specifically, we solve hydrodynamics equations within the edge
length of $\sim 1500$ km, and thus we can track ejecta motion up to
$\sim 10$ ms taking the fact that the ejecta velocity is usually smaller
than $0.5 c$, where $c$ is the speed of light. Asymptotic properties
such as the velocity and kinetic energy would not be correctly estimated
if we measure them in a near zone. Hence, the large computational domain
is essential for an accurate study of the ejecta. Second, we decrease
the density of artificial atmosphere, which is inevitable in
conservative hydrodynamics schemes. The atmospheric density is at most
$10^3 \; \mathrm{g \; cm^{-3}}$, and negligible for ejecta (see
Fig.~\ref{fig:ejmer}). Indeed, we confirmed that ejecta properties
depend very weakly on the atmospheric density as far as it is
sufficiently low. Specifically, varying the atmospheric density by an
order of magnitude changes the ejecta properties only by
$10\%$-$20\%$. Finally, we improve grid resolutions by $\sim 20\%$ so
that the NS radius is covered by $\approx 50$ grid points in the highest
resolution runs. The radius of the ejecta is always covered by $\approx
50$-60 grid points in the equatorial plane, and $\approx 10$ grid points
in the perpendicular direction (see the next section). We perform
simulations with 3 different resolutions for selected models, and
estimate that ejecta properties are accurate within $\approx 10\%$ for
many cases and within a factor of 2 for the worst cases in which the
ejecta mass is small. This accuracy is sufficient for the purpose of
this article, which mainly discuss qualitative signatures. A convergence
study will be presented in a separate paper with detailed discussions of
our systematic simulations.

\begin{table}[tbp]
 \caption{Important values of representative models measured at 10 ms
 after the onset of the merger for highest resolution
 runs. $M_\mathrm{disk}$ and $M_\mathrm{ej}$ are the bound and unbound
 masses, respectively, outside the apparent horizon of the
 BH. $T_\mathrm{ej}$ is the kinetic energy of the ejecta, and
 $v_\mathrm{ej}$ is the bulk velocity of the ejecta.}
 \begin{tabular}{ccc|cccc} \hline
  $Q$ & $\chi$ & EOS & $M_\mathrm{disk} [M_\odot]$ & $M_\mathrm{ej}
  [M_\odot]$ & $T_\mathrm{ej}$ (erg) & $v_\mathrm{ej} [c]$ \\
  \hline \hline
  3 & 0.75 & APR4 & 0.18 & 0.01 & $5 \times 10^{50}$ & 0.19 \\
  3 & 0.75 & ALF2 & 0.23 & 0.05 & $3 \times 10^{51}$ & 0.21 \\
  3 & 0.75 & H4 & 0.29 & 0.05 & $2 \times 10^{51}$ & 0.20 \\
  3 & 0.75 & MS1 & 0.29 & 0.07 & $4 \times 10^{51}$ & 0.21 \\
  3 & 0 & MS1 & 0.14 & 0.02 & $8 \times 10^{50}$ & 0.19 \\
  5 & 0.75 & H4 & 0.27 & 0.05 & $3 \times 10^{51}$ & 0.22 \\
  7 & 0.75 & H4 & 0.16 & 0.04 & $3 \times 10^{51}$ & 0.19 \\
  \hline
 \end{tabular}
 \label{table:result}
\end{table}

\section{Mass ejection}

When the NS is disrupted by the BH tidal field, a one-armed spiral
structure called the tidal tail is formed around the BH. Although a
large part of the tail eventually falls back onto the remnant disk and
BH, its outermost part obtains a sufficient angular momentum and kinetic
energy to become unbound via hydrodynamic angular momentum transport
processes. Dynamical mass ejection from the BH-NS merger is driven
dominantly by this tidal effect. We also find that some material in the
vicinity of the BH becomes unbound when the tidal tail hits itself as it
spirals around the BH. This ejection may be ascribed to the shock
heating, but this shock-driven component is always subdominant.

Ejecta exhibits a crescentlike shape as depicted in Fig.~\ref{fig:ejmer}
in most cases. Specifically, a typical opening angle of the ejecta in
the equatorial plane is $\varphi_\mathrm{ej} \approx \pi$. Such a
nonaxisymmetric shape arises because the sound-crossing time scale is
shorter than the orbital period at the onset of tidal
disruption. Furthermore, ejecta spreads dominantly in the equatorial
plane, and expands only slowly in the direction perpendicular to the
equatorial plane (hereafter, the {\it z}-direction). The reason for this
is that the ejection is driven mainly by the tidal effect, which is most
efficient in the equatorial plane. Thus, a portion of circumferential
material will be subsequently swept by the ejecta. A typical half
opening angle of the ejecta around the equatorial plane is
$\theta_\mathrm{ej} \approx 1/5$ radian; and, this implies that the
ejecta velocity in the equatorial plane $v_\parallel$ is larger by a
factor of $1 / \theta_\mathrm{ej} \approx 5$ than that in the {\it
z}-direction $v_\perp$. Here, $v_\parallel$ may be identified with the
radial velocity, and the azimuthal velocity should become negligible
soon after the ejection due to the angular momentum
conservation. Indeed, azimuthal velocity is very small in
Fig.~\ref{fig:ejmer}. Aside from the ejecta itself, the region above the
remnant BH is much clearer than that for the NS-NS merger, and thus the
baryon-loading problem of GRB jets may be less severe.

The ejecta mass $M_\mathrm{ej}$ depends on binary parameters and NS EOSs
and are typically in the range of $\sim 0.01$--$0.1 M_\odot$ when the
tidal disruption occurs and the disk mass $M_\mathrm{disk}$ exceeds
$\sim 0.1 M_\odot$. Important values are shown in
Table~\ref{table:result} for representative models in which
$M_\mathrm{disk} \gtrsim 0.1 M_\odot$. The value of $M_\mathrm{ej}$ is
generally large when the NS EOS is stiff and the NS radius is large for
fixed values of $Q$ and $\chi$, because the mass ejection is driven
primarily by the tidal effect. This dependence on the NS EOS and radius
is opposite to the case of the NS-NS merger in general relativity, where
rapid rotation and oscillation of a remnant massive NS drive ejection
\cite{hotokezaka_kkosst2013,bauswein_gj2013}. We speculate that ejecta
from the BH-NS merger might account for a substantial portion of the
{\it r}-process nucleosynthesis (see also below) compared to that from
the NS-NS merger if the realistic NS EOS is stiff, and vice versa. When
the NS is not disrupted prior to the merger, the ejecta mass is
negligible for current astrophysical interest. The ejecta mass is always
smaller than the disk mass, and a relation $M_\mathrm{ej} \approx
0.05$--$0.25 M_\mathrm{disk}$ approximately holds for a wide range. When
the disk and ejecta are very massive, it seems that $M_\mathrm{ej}$ can
be a larger fraction of $M_\mathrm{disk}$ than this relation indicates.

The ejecta from the BH-NS merger has a bulk linear momentum,
$P_\mathrm{ej}$, and resulting large bulk velocity $v_\mathrm{ej} \equiv
P_\mathrm{ej} / M_\mathrm{ej} \sim 0.2 c$, in a particular
direction. These bulk linear momentum and velocity would essentially
vanish for nearly spherical ejecta such as one from the NS-NS
merger. The value of $v_\mathrm{ej} \sim 0.2 c$ depends only weakly on
the binary parameters and NS EOS as far as the mass ejection is
substantial.

Typical values of kinetic energy $T_\mathrm{ej}$ are in the range $\sim
5 \times 10^{50}$--$5 \times 10^{51}$ erg. The average velocity of the
ejecta $v_\mathrm{ave} \approx ( 2 T_\mathrm{ej} / M_\mathrm{ej}
)^{1/2}$ is typically 0.25--$0.3 c$ and is naturally larger than
$v_\mathrm{ej}$. The contribution of $v_\perp$ to $T_\mathrm{ej}$ is
smaller by a factor of $\theta_\mathrm{ej}^2$ than that of
$v_\parallel$, and thus a relation $v_\mathrm{ave} \approx v_\parallel$
holds. For an axisymmetric outflow truncated at an opening angle
$\varphi_\mathrm{ej}$, a relation $v_\mathrm{ej} / v_\parallel \approx
\sin (\varphi_\mathrm{ej} /2) / ( \varphi_\mathrm{ej} /2)$ should hold,
and thus the ejecta opening angle in the equatorial plane is estimated
to be $\varphi_\mathrm{ej} \approx (0.7$--$1.3) \pi$ radian. This is
consistent with Fig.~\ref{fig:ejmer}.

\section{Discussion}

In this section, we discuss possible consequences of the anisotropic
ejecta from the BH-NS merger in chronological order as closely as
possible. For convenience, we introduce $M_{1.35} \equiv M_\mathrm{NS} /
1.35 M_\odot$, $A_5 \equiv (1+Q)/(1+5)$, $M_\mathrm{ej,0.03} \equiv
M_\mathrm{ej} / 0.03 M_\odot$, $v_{\mathrm{ej},0.2} \equiv v_\mathrm{ej}
/ 0.2 c$, $v_{\mathrm{ave},0.3} \equiv v_\mathrm{ave} / 0.3 c$,
$\theta_{\mathrm{ej},i} \equiv \theta_\mathrm{ej} / (1/5)$, and
$\varphi_{\mathrm{ej},i} \equiv \varphi_\mathrm{ej} / \pi$.

\subsection{Kick velocity of the black hole}

The remnant BH should receive backreaction from the ejecta with
$M_\mathrm{ej}$ and $v_\mathrm{ej}$, and obtain substantial ``ejecta
kick'' velocity \cite{rosswog_dtp2000}. The total mass of the remnant BH
and surrounding disk is determined approximately by the mass of the
binary at infinite separation $m_0 = (1+Q) M_\mathrm{NS}$, neglecting
$M_\mathrm{ej}$ and energy carried by GWs $\approx 0.05 m_0 c^2$. Thus,
the ejecta kick velocity of the remnant BH-disk will be
\begin{equation}
 v_\mathrm{kick} \approx \frac{M_\mathrm{ej}}{m_0} v_\mathrm{ej} = 220
  \; \mathrm{km \; s^{-1}} M_{\mathrm{ej},0.03} v_{\mathrm{ej},0.2}
  M_{1.35}^{-1} A_5^{-1} .
\end{equation}
This value is larger than escape velocity of globular clusters and dwarf
galaxies for many cases, and can exceed that of small galaxies under
suitable conditions \cite{merritt_mfhh2004} .

The ejecta kick velocity, $v_\mathrm{kick}$, can be larger than the kick
velocity due to GW radiation reaction. Specifically, the GW kick
velocity is at most $\sim 150 \; \mathrm{km \; s^{-1}}$ when the NS is
disrupted prior to the merger, even if the BH spin is misaligned
\cite{shibata_kyt2009,kyutoku_ost2011,foucart_dkt2011,foucart_ddkmopsst2013}. The
reason for this is that the linear momentum is radiated most efficiently
when the binary is about to merge, and, therefore, earlier disruption
significantly suppresses the GW kick velocity. Thus, the ejecta kick
velocity can dominate the velocity of the remnant when tidal disruption
is prominent.

\subsection{Gravitational-wave memory}

Because the ejecta and remnant BH-disk travel in the opposite direction
after the merger, nonoscillatory GW emission, i.e., GW memory, is
expected. Assuming that the remnant mass $\approx m_0$ is much larger
than $M_\mathrm{ej}$, magnitude of this (linear) ejecta memory is given
approximately by \cite{braginsky_thorne1987}
\begin{equation}
 \delta h \approx \frac{2 G M_\mathrm{ej} v_\mathrm{ej}^2}{c^4 D} = 1.1
  \times 10^{-24} M_{\mathrm{ej},0.03} v_\mathrm{ej,0.2}^2 D_2^{-1} ,
\end{equation}
where $G$ is the gravitational constant and $D \equiv D_2 \times 100$
Mpc is a distance from the binary to the observer. Taking the fact that
the expected rise time of ejecta memory is much shorter than the inverse
of frequency at which ground-based detectors are most sensitive, $( \sim
100 \; \mathrm{Hz} )^{-1}$, it may be possible to detect ejecta memory
by the Einstein Telescope \cite{et2012} if the ejecta is as massive as
$\gtrsim 0.1 M_\odot$. A possibility of such significant mass ejection
is not negligible if the BH spin can be very large
\cite{lovelace_dfkpss2013}.

The ejecta memory as well as GWs from the binary coalescence can be
strongest in the {\it z}-direction. Such memory would easily be
distinguished from linear memory from GRB jets \cite{sago_iny2004} and
nonlinear memory from coalescence GWs \cite{favata2010}, which are very
weak in the {\it z}-direction, if we observe the binary from this
direction.

\subsection{Macronova/kilonova}

The macronova a.k.a. kilonova is quasithermal radiation from the ejecta
heated by the radioactive decay of {\it r}-process elements
\cite{li_paczynski1998,kulkarni2005,metzger_mdqaktnpz2010}. Although
properties of {\it r}-process elements such as opacities are still
uncertain so that accurate predictions are difficult
\cite{barnes_kasen2013,kasen_bb2013}, the macronova/kilonova is one of
the most promising EM counterparts to the BH-NS merger.

Peak values of the macronova/kilonova are estimated when the diffusion
time scale becomes equal to the dynamical time scale, which is
essentially the time after the merger. Assuming random walks of photons
in spherical ejecta, the peak time is estimated to be $t_\mathrm{peak,s}
\approx ( 3 \kappa M_\mathrm{ej} / 4 \pi c v_\mathrm{ave} )^{1/2} = 8 \;
\mathrm{day} \; \kappa_1^{1/2} M_{\mathrm{ej},0.03}^{1/2}
v_{\mathrm{ave},0.3}^{-1/2}$, where $\kappa \equiv \kappa_1 \times 10 \;
\mathrm{g^{-1} \; cm^2}$ is the opacity. If a fraction $f \equiv f_{-6}
\times 10^{-6}$ of $M_\mathrm{ej} c^2$ is radiated, the peak luminosity
is expected to be $L_\mathrm{peak,s} \approx f M_\mathrm{ej} c^2 /
t_\mathrm{peak,s} = 7 \times 10^{40} \; \mathrm{erg \; s^{-1}} \; f_{-6}
\kappa_1^{-1/2} M_{\mathrm{ej},0.03}^{1/2} v_{\mathrm{ej},0.3}^{1/2}$,
and the temperature at the peak is $T_\mathrm{peak,s} \approx (
L_\mathrm{peak,s} / 4 \pi v_\mathrm{ave}^2 t_\mathrm{peak,s}^2 \sigma
)^{1/4} = 1200 \; \mathrm{K} \; f_{-6}^{1/4} \kappa_1^{-3/8}
M_{\mathrm{ej},0.03}^{-1/8} v_{\mathrm{ej},0.3}^{-1/8}$, where $\sigma$
is the Stefan-Boltzmann constant.

The emission could be modified by ejecta geometry for the BH-NS
merger. We approximate ejecta geometry by an axisymmetric cylinder with
$v_\parallel$ and $v_\perp$ in the radial and {\it z}-directions,
respectively, and truncate at an opening angle
$\varphi_\mathrm{ej}$. Assuming that photons escape from the {\it
z}-direction due to a short distance, the peak time is estimated as
\begin{align}
 t_\mathrm{peak} & \approx \left( \frac{\kappa M_\mathrm{ej}
 v_\perp}{c \varphi_\mathrm{ej} v_\parallel^2} \right)^{1/2} \notag \\
 & = 4 \; \mathrm{day} \; \kappa_1^{1/2} M_{\mathrm{ej},0.03}^{1/2}
 v_{\mathrm{ave},0.3}^{-1/2} \theta_{\mathrm{ej},i}^{1/2}
 \varphi_{\mathrm{ej},i}^{-1/2} .
\end{align}
Accordingly, the peak luminosity is
\begin{align}
 L_\mathrm{peak} & \approx \frac{f M_\mathrm{ej} c^2}{t_\mathrm{peak}}
 \notag \\
 & = 1.4 \times 10^{41} \; \mathrm{erg} \; f_{-6} \kappa_1^{-1/2}
 M_{\mathrm{ej},0.03}^{1/2} v_{\mathrm{ej},0.3}^{1/2}
 \theta_{\mathrm{ej},i}^{-1/2} \varphi_{\mathrm{ej},i}^{1/2} ,
\end{align}
and finally the temperature at the peak is
\begin{align}
 T_\mathrm{peak} & \approx \left(
 \frac{L_\mathrm{peak}}{\varphi_\mathrm{ej} v_\parallel^2
 t_\mathrm{peak}^2 \sigma} \right)^{1/4} \notag \\
 & = 3000 \; \mathrm{K} \; f_{-6}^{1/4} \kappa_1^{-3/8}
 M_{\mathrm{ej},0.03}^{-1/8} v_{\mathrm{ave},0.3}^{-1/8}
 \theta_{\mathrm{ej},i}^{-3/8} \varphi_{\mathrm{ej},i}^{1/8} .
\end{align}
These rough estimates suggest that characteristics of the
macronova/kilonova associated with anisotropic ejecta from the BH-NS
merger will be modified by a factor of a few compared to those
associated with the isotropic one from the NS-NS merger when the values
of the ejecta mass and velocity are the same.

Directional dependence is important for the macronova/kilonova of the
BH-NS merger \cite{tanaka_etal2012}. In particular, the flux will differ
by a factor of $\approx 1/\theta_\mathrm{ej} \approx 5$ for different
viewing angles at $t_\mathrm{peak}$. Specifically, the radiation will be
dim if we are located in the equatorial plane. The time evolution of
light curves will also differ, because the diffusion timescale is
different for different viewing angles. Although atomic line structures
of {\it r}-process elements will be blended due to the variety of
elements and energy levels, observation might be possible from the {\it
z} direction, for which the line broadening is not severe due to the
small velocity, $v_\perp$, if some prominent lines are
isolated. Blueshifts of lines are also small in the {\it z}
direction. In addition, linear polarization up to $\approx 4\%$--$5\%$
is expected for an aspherical photosphere observed in the equatorial
plane \cite{Hoflich1991}.

\subsection{Synchrotron radio flare}

Blast waves will develop between the ejecta and interstellar medium
(ISM) as the ejecta travels in the ISM. If electrons are accelerated to
nonthermal velocity distribution and magnetic fields are amplified
behind the blast waves in a similar manner to supernova remnants and GRB
afterglows, synchrotron radio flares should arise
\cite{nakar_piran2011}. The synchrotron radio flare is also one of the
most promising EM counterparts.

The peak time of the flare is the time at which the ejecta accumulates
the mass comparable to its own from the ISM with density $n_\mathrm{H}
\equiv n_{\mathrm{H},0} \times 1 \; \mathrm{cm}^{-3}$ and begins to be
decelerated. The deceleration radius is given by $R_\mathrm{dec,s} = ( 3
M_\mathrm{ej} / 4 \pi m_\mathrm{p} n_\mathrm{H} )^{1/3} = 0.7 \;
\mathrm{pc} \; M_{\mathrm{ej},0.03}^{1/3} n_{\mathrm{H},0}^{-1/3}$ for
spherical ejecta, and thus the deceleration time is given by
$t_\mathrm{dec,s} = R_\mathrm{dec,s} / v_\mathrm{ave} = 7 \; \mathrm{yr}
\; M_{\mathrm{ej},0.03}^{1/3} n_{\mathrm{H},0}^{-1/3}
v_{\mathrm{ave},0.3}^{-1}$.

The ejecta from the BH-NS merger is not spherical, and can accumulate
only a small portion of the ISM within a fixed radius. Thus, the ejecta
has to travel a longer distance $R_\mathrm{dec} > R_\mathrm{dec,s}$ to
be decelerated than spherical ejecta. Here, we assume that values of
$v_\parallel$ and $v_\perp$ do not change significantly before the
deceleration. This would lead to constant values of $\theta_\mathrm{ej}$
and $\varphi_\mathrm{ej}$ before the deceleration, which is expected to
occur at
\begin{align}
 R_\mathrm{dec} & \approx R_\mathrm{dec,s} \left( \frac{\pi /
 2}{\theta_\mathrm{ej}} \right)^{1/3} \left( \frac{2
 \pi}{\varphi_\mathrm{ej}} \right)^{1/3} \notag \\
 & = 1.7 \; \mathrm{pc} \; M_{\mathrm{ej},0.03}^{1/3}
 n_{\mathrm{H},0}^{-1/3} \theta_{\mathrm{ej},i}^{-1/3}
 \varphi_{\mathrm{ej},i}^{-1/3} ,
\end{align}
and, therefore, the peak time is
\begin{equation}
 t_\mathrm{dec} = \frac{R_\mathrm{dec}}{v_\parallel} = 18 \; \mathrm{yr}
  \; M_{\mathrm{ej},0.03}^{1/3} v_{\mathrm{ave},0.3}^{-1}
  n_{\mathrm{H},0}^{-1/3} \theta_{\mathrm{ej},i}^{-1/3}
  \varphi_{\mathrm{ej},i}^{-1/3} .
\end{equation}
The peak time should be closer to (but still longer than due to the
energy conservation) $t_\mathrm{dec,s}$ if the ejecta approaches a
spherical state. The reality may exist between these two limits. We left
precise modelings of geometry evolution for the future study.

The peak luminosity and flux of the flare do not depend on geometry as
far as ejecta evolves adiabatically. The reason for this is that the
peak time is still the time at which the ejecta accumulate
$M_\mathrm{ej}$ from the ISM, and thus the number of emitting electrons
and their characteristic frequency do not change. Directional dependence
of the flare as well as that of the macronova/kilonova is worth
studying.

\subsection{Proper motion of radio images}

Because the ejecta moves with $v_\mathrm{ej}$, radio observation will be
able to detect the proper motion of radio images in the timescale of
$t_\mathrm{dec}$ in addition to expansion of images. A characteristic
travel distance may be given by $R_\mathrm{dec} ( v_\mathrm{ej} /
v_\mathrm{ave} )$. Averaging over random distribution of ejecta
direction halves the projected distance. In reality, however, GW
observation may be biased toward the {\it z} axis, and then the biased
average could be larger. We expect the projected proper motion of radio
images to be $O(1)$ pc around the synchrotron radio flare peak. This
implies that the radio image is expected to move $O(1)$ mas for a BH-NS
binary at $O(100)$ Mpc. Current radio instruments may resolve this angle
\cite{taylor_fbk2003}.

The proper motion will give us a way to distinguish the BH-NS merger
from the NS-NS merger, because proper motion of radio images should be
much smaller for the latter, whereas expansion of images should be
common to two merger types. It is not easy to distinguish between BH-NS
and NS-NS binaries only from GW observation, especially when the BH is
less massive \cite{hannam_bffh2013}. Thus, information provided by EM
counterparts is also useful for the determination of merger types
\cite{stone_lb2013,kyutoku_is}.

\subsection{Cosmic-ray acceleration}

Physical condition of the ejecta is expected to be similar to that of a
supernova remnant. This implies that the cosmic-ray (CR) acceleration
will occur in the remnant of ejecta from the BH-NS merger via diffusive
shock acceleration \cite{bell1978,blandford_ostriker1978}, as well as
the NS-NS merger. The maximum attainable energy of CRs is given by the
Hillas condition $\varepsilon_\mathrm{max} = Z e \beta B R$
\cite{hillas1984}, where $Z$, $\beta$, $B$, and $R$ are the proton
number, shock velocity divided by $c$, magnetic field, and typical
source size, respectively. Assuming that a fraction $\epsilon_B \equiv
\epsilon_{B,-1} \times 0.1$ of $T_\mathrm{ej}$ is converted to magnetic
field energy, we obtain
\begin{equation}
 \varepsilon_\mathrm{max} = 2.4 \times 10^{18} \; \mathrm{eV} \; Z
  M_{\mathrm{ej},0.03}^{1/3} v_{\mathrm{ave},0.3}^2
  \epsilon_{\mathrm{B},-1}^{1/2} n_{\mathrm{H},0}^{1/6}
  \theta_{\mathrm{ej},i}^{2/3} \varphi_{\mathrm{ej},i}^{-1/3} ,
\end{equation}
where we adopt $R = R_\mathrm{dec} \theta_\mathrm{ej}$. The higher
energy than that for supernova remnants primarily owes to the faster
ejecta velocity.

Cosmic rays from the galactic BH-NS merger could explain observed CRs
above the knee region up to the ankle region. Taking the fact that
ejecta kinetic energy is comparable to that of supernova remnants and
comparing galactic BH-NS merger rate estimation $\sim 10^{-7}$--$10^{-4}
\; \mathrm{yr}^{-1}$ \cite{ligovirgo2010} with the galactic supernova
rate $\sim 0.01 \; \mathrm{yr}^{-1}$, the CR energy above the knee can
be explained. Evidently, acceleration, propagation, and galactic
confinement of CRs including nuclei have to be investigated in more
detail.

Acceleration of heavy elements like irons to even an ultrahigh energy
$\gtrsim 10^{20}$ eV are also possible for massive ejecta $\gtrsim 0.2
M_\odot$ such as the ones reported in
\cite{lovelace_dfkpss2013}. Although an energetics constraint is not
assuring, observed energy of ultrahigh energy CRs
\cite{murase_takami2009} is marginally consistent within the uncertainty
of merger rate estimation. Assuming that a fraction
$\epsilon_\mathrm{CR} \equiv \epsilon_{\mathrm{CR},-1} \times 0.1$ of
ejecta kinetic energy averaged over BH-NS binary distribution $\langle
T_\mathrm{ej} \rangle \equiv \langle T_{\mathrm{ej},51} \rangle \times
10^{51}$ erg is converted to CR energy, the required merger rate is
approximately $\epsilon_{\mathrm{CR},-1} \langle T_\mathrm{ej,51}
\rangle^{-1} \; \mathrm{Mpc^{-3} \; Myr^{-1}}$, where $1 \;
\mathrm{Mpc}^{-3} \; \mathrm{Myr}^{-1}$ is the ``high'' estimation
\cite{ligovirgo2010}.

\begin{acknowledgments}
 We are grateful to Alan G. Wiseman, Masaomi Tanaka, and Kenta
 Hotokezaka for valuable discussions. This work was supported by the
 Grant-in-Aid for Scientific Research Grants No.~21684014, No.~22244030,
 No.~24000004, No.~24103006, and No.~24740163 of Japanese MEXT, and by
 JSPS Postdoctoral Fellowship for Research Abroad.
\end{acknowledgments}

\end{document}